\documentclass[twocolumn]{aastex701}

\newcommand\multimoon{\texttt{MultiMoon}}
\newcommand\jt{$J_2$}
\newcommand\ct{$C_{22}$}

\usepackage[utf8]{inputenc}
\usepackage{tipa}
\usepackage{combelow}
\usepackage{savesym}
\savesymbol{tablenum}
\usepackage{siunitx}
\restoresymbol{SIX}{tablenum}
\usepackage{breqn}
\usepackage{tabularx}
\usepackage{amsmath}
\usepackage{paralist}
\usepackage{graphicx}
\usepackage{subcaption}
\usepackage{placeins}
\usepackage{comment}

\begin{document}

\title{The Dense, Rocky Core of Haumea Revealed by Satellite Dynamics}

\author[orcid=0000-0002-1788-870X,sname='Proudfoot']{Benjamin Proudfoot}
\affiliation{Florida Space Institute, University of Central Florida, 12354 Research Parkway, Orlando, FL 32826, USA}
\email[show]{benp175@gmail.com}

\author[orcid=0000-0003-1080-9770, sname='Ragozzine']{Darin Ragozzine} 
\affiliation{Brigham Young University Department of Physics \& Astronomy, N283 ESC, Brigham Young University, Provo, UT 84602, USA}
\email{darin_ragozzine@byu.edu}

\author[orcid=0000-0002-8296-6540, sname='Grundy']{Will Grundy} 
\affiliation{Northern Arizona University, Department of Astronomy \& Planetary Science, PO Box 6010, Flagstaff, AZ 86011, USA}
\email{}

\author[orcid=0000-0002-6085-3182,sname='Rommel']{Flavia L. Rommel}
\affiliation{Florida Space Institute, University of Central Florida, 12354 Research Parkway, Orlando, FL 32826, USA}
\email{flavialuane.rommel@ucf.edu}

\author[orcid=0000-0002-6117-0164]{Bryan J. Holler}
\affiliation{Space Telescope Science Institute, Steven Muller Building, 3700 San Martin Drive Baltimore, MD 21218 Baltimore, MD, USA}
\email{bholler@stsci.edu}

\begin{abstract}

The icy dwarf planets in the outer solar system are among the most enigmatic bodies known, showing activity ranging from atmospheric resurfacing to possible cryovolcanism. Yet the study of their interiors remains largely theoretical, with few empirical measurements to guide our understanding. \added{The orbits of their moons} offer a way to break this deadlock, providing a direct probe of dwarf planet interiors. Here we present a detailed investigation of Haumea's interior derived from the motion of its satellite system. Fitting a new dynamical model to two decades of precise \textit{Hubble Space Telescope} astrometry, we simultaneously measure Haumea's dynamical oblateness, $J_2$, and the masses of its satellites. Our analysis of the \jt{} measurement combined with occultation-derived shape models conclusively shows that Haumea has a dense rocky core overlaid with an ice-rich mantle. However, the current dataset is not constraining enough to fully determine the detailed internal structure (core/mantle density, two-layer vs. three-layer, etc.). We do find a slight preference for models where Haumea retains a frozen-in fossil figure, but more data are needed to statistically confirm this. Future observations of the satellite system, as well as observations of stellar occultations, are needed to fully test this hypothesis and should substantially tighten constraints on Haumea's interior. Taken together with other recent evidence for geophysical and geochemical evolution of dwarf planets, our work shows that dense rocky cores and subsurface oceans are likely to be ubiquitous among the trans-Neptunian dwarf planets.

\end{abstract}


\section{Introduction}
Alongside the solar system's planets, dwarf planets provide an additional source of evidence for our understanding of the planet formation process. \added{These embryonic planetary relicts were traditionally understood as cold, dead worlds based on expectations that their small sizes would preclude sufficient internal heat retention to drive any geological activity.} However, in the past decades, with improved remote observations and robotic exploration \added{of Pluto and Charon} by the \textit{New Horizons} spacecraft, these bodies have been revealed as dynamic, active worlds \added{\citep[e.g,][]{2015Sci...350.1815S,BEYER2017161,DESCH2017175,2018ARA&A..56..357S,moore2021geologically,singer2025pluto}}. One particular focus of geophysical interest is the interior structure of these bodies, specifically their ability to form and sustain long-lived liquid water subsurface oceans \citep{desch2009thermal}. Various indirect lines of evidence --- like tidal modeling \citep{nimmo2023internal,akiba2026eris}, analysis of isotopic ratios \citep{grundy2024measurement,glein2024moderate}, and even possible detections of active hotspots \citep{kiss2024prominent} --- suggest that such subsurface oceans did once, and may still, exist \added{on some of these bodies}. \added{Subsurface oceans are of potential astrobiological interest. Water-rock interfaces in these bodies, whether in subsurface oceans \citep{deamer2017can} or within liquid water inside their cores \citep{dunham2019haumea}}, may resemble hydrothermal settings on Earth one of the candidate locations for the emergence of life \citep[e.g.,][]{baross1985submarine}. Despite \added{the possible astrobiological importance of these worlds}, precious few direct constraints can be definitively placed on the interiors of distant, icy dwarf planets. 


\added{Compared with the population of dwarf planets and trans-Neptunian objects (TNOs), Haumea stands out as a particularly peculiar world. With a rotation period of $\sim$3.9 hours \citep[e.g.,][]{rabinowitz2006photometric}, Haumea has an extreme shape dominated by rotational forces. Haumea's surface is unique, showing the strong spectral features of a nearly pure water ice surface \citep{2007Nature..446..296,2007ApJ...655.1172T,fraser2009nicmos,2009A&A...496..547P,dumas2011high}.
Surrounding Haumea is a ring dynamically confined by Haumea's gravitational field \citep{ortiz2017size}, as well as the two known satellites. Haumea is also host to the trans-Neptunian region's only known collisional family, identified by dynamical proximity to Haumea and the same unique spectral features as Haumea itself \citep{2007Nature..446..296,2007AJ....134.2160R,2008ApJ...684L.107S,proudfoot2019modeling,2024AJ....168..269P}. Explaining the origin of Haumea
is difficult \citep[see][]{proudfoot2019modeling}, but \added{additional constraints on} its interior may provide clues.}

\added{Among the dwarf planets in the trans-Neptunian region, Haumea is the best candidate for a detailed investigation of its interior \citep{proudfoot2024bpm3}. Haumea has two known satellites, Namaka and Hi'iaka \citep{brown2005keck,brown2006satellites}, that have both strong perturbations from mutual gravitational interactions \citep[][]{rb09} --- allowing for direct measurement of the satellite masses --- and from Haumea's nonspherical shape \citep{rabinowitz2006photometric,ortiz2017size}. Dynamical simulations show that it is possible to independently measure the satellite masses as well as Haumea's gravitational harmonics, though the data have not yet been sufficient for breaking degeneracies in these parameters \citep{proudfoot2024bpm3}. As with other worlds, since Haumea's external shape has been precisely measured by stellar occultations \citep{ortiz2017size,proudfoot2026occ}, these gravitational harmonics directly probe the density distribution within Haumea. }

\added{Although Haumea is unique in many regards, it still operates under the same physics that dominate the other dwarf planets. Melting and differentiation driven by internal heating from long-lived radioisotopes, impacts and collisions creating dynamic ring and satellite systems, and geochemical evolution through hydration reactions, are all processes that appear to have affected the diverse population of dwarf planets. Thus, studying these processes for Haumea, allows us to learn about the range of evolution that other dwarf planets have undergone.}

\begin{figure*}
    \centering\includegraphics[width=0.99\textwidth]{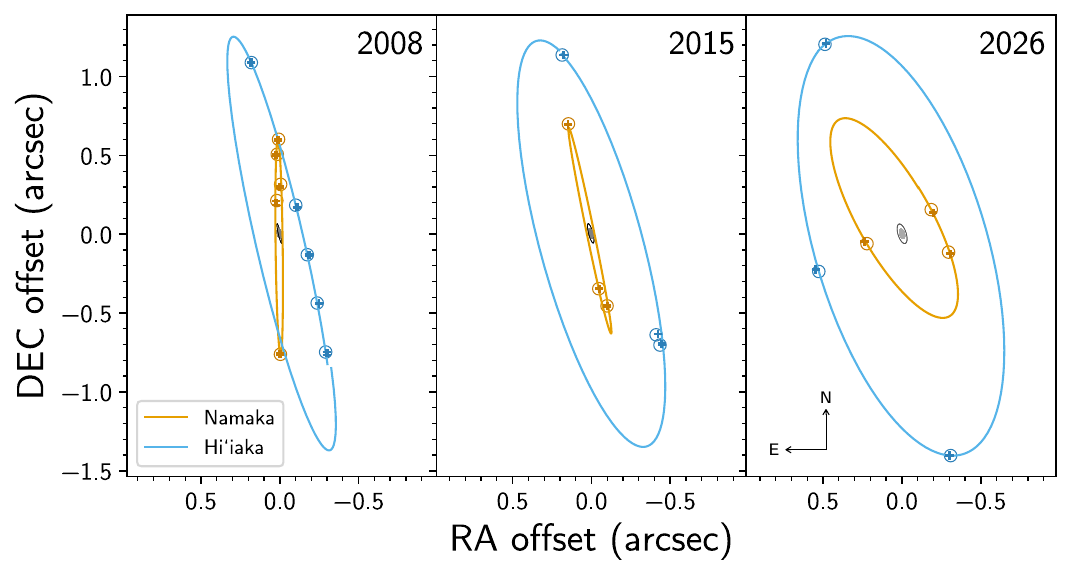}
    \caption{Our orbit solution projected on the sky at the three epochs for which there are multi-day observational baselines. Solid lines show the orbit solutions for Namaka (orange) and Hi`iaka (blue) over a single orbit at the beginning of each epoch compared with the measured astrometry (darker points with error bars). Open circles show the model predicted position at the time of each detection. Haumea and its ring, shown at roughly the correct scale and orientation, are shown at the origin. }
    \label{fig:orbit}
\end{figure*}

\begin{deluxetable}{lcc}
\tablecaption{Haumea System Orbit Solution\label{tab:orbitfit}}
\tablewidth{0pt}
\tablehead{
\colhead{Parameter} & \colhead{Symbol} & \colhead{Posterior}
}
\startdata
\multicolumn{3}{c}{Haumea} \\
Mass ($10^{18}$ kg)                & $M_{P}$         & $3904^{+7}_{-7}$             \\
$J_2$ harmonic                     & $J_2$           & $0.143^{+0.025}_{-0.024}$ \\
Rotation axis obliquity ($\degr$)  & $i_{sp}$        & $77.0^{+0.4}_{-0.5}$           \\
Rotation axis precession ($\degr$) & $\Omega_{sp}$   & $15.0^{+0.6}_{-0.7}$           \\
\hline
\multicolumn{3}{c}{Namaka} \\
Mass ($10^{18}$ kg)                & $M_N$           & $1.35^{+0.08}_{-0.08}$       \\
Semi-major axis (km)               & $a_N$           & $25365^{+16}_{-16}$          \\
Eccentricity                       & $e_N$           & $0.220^{+0.002}_{-0.002}$ \\
Inclination ($\degr$)              & $i_N$           & $68.9^{+0.1}_{-0.1}$        \\
Argument of periapse($\degr$)      & $\omega_N$      & $118.3^{+0.3}_{-0.3}$        \\
Nodal longitude ($\degr$)          & $\Omega_N$      & $23.90^{+0.13}_{-0.13}$   \\
Mean anomaly ($\degr$)             & $\mathcal{M}_N$ & $184.7^{+0.4}_{-0.4}$        \\
\hline
\multicolumn{3}{c}{Hi`iaka} \\
Mass ($10^{18}$ kg)                & $M_H$           & $16.5^{+0.5}_{-0.5}$       \\
Semi-major axis (km)               & $a_H$           & $49186^{+27}_{-27}$          \\
Eccentricity                       & $e_H$           & $0.0604^{+0.0008}_{-0.0008}$ \\
Inclination ($\degr$)              & $i_H$           & $77.24^{+0.04}_{-0.04}$      \\
Argument of periapse($\degr$)      & $\omega_H$      & $108.4^{+0.7}_{-0.7}$        \\
Nodal longitude ($\degr$)          & $\Omega_H$      & $13.09^{+0.03}_{-0.03}$   \\
Mean anomaly ($\degr$)             & $\mathcal{M}_H$ & $144.5^{+0.7}_{-0.7}$        \\
\hline
\multicolumn{3}{c}{Derived Parameters} \\
Namaka inclination ($\degr$)       & $\varepsilon_N$ & $11.8^{+0.5}_{-0.5}$         \\
Hi`iaka inclination ($\degr$)      & $\varepsilon_H$ & $1.9^{+0.6}_{-0.6}$          \\
Haumea pole R.A. ($\degr$)         & $\alpha_{\rm rot}$  & $284.8^{+0.6}_{-0.7}$         \\
Haumea pole dec. ($\degr$)         & $\delta_{\rm rot}$  & $-9.7^{+0.5}_{-0.4}$         \\
Namaka orbit pole R.A. ($\degr$)   & $\alpha_{N}$    &   $292.21^{+0.11}_{-0.11}$  \\
Namaka orbit pole dec. ($\degr$)   & $\delta_{N}$    &   $-0.48^{+0.11}_{-0.11}$    \\
Hi`iaka orbit pole R.A. ($\degr$)  & $\alpha_{H}$    &  $282.97^{+0.03}_{-0.03}$      \\
Hi`iaka orbit pole dec. ($\degr$)  & $\delta_{H}$    &  $-10.09^{+0.04}_{-0.04}$        \\
\enddata
\tablecomments{Reported values for parameter posterior distributions show the median, 16th, and 84th percentile values. All fitted angles are relative to the J2000 ecliptic plane and referenced to Haumea-centric JD 2454615.0 (2008 May 28 12:00 UT). R.A. and decl. values are relative to the J2000 equatorial plane. Haumea reference radius taken to be 772 km \citep{proudfoot2026occ}.}
\end{deluxetable}

\section{The Haumea System's Orbital Configuration}
In this work, we leverage three new epochs of imaging from the \textit{Hubble Space Telescope} (HST) \added{in an effort to determine Haumea's interior structure by measuring Haumea's gravitational harmonics.} When combined with an archive of HST observations taken since 2006, we have been able to fit a new dynamical model to the Haumea system (see Figure \ref{fig:orbit} and Table \ref{tab:orbitfit}). For further details on our observations and orbit fitting methods, which are applications of previously published methods, see Appendices \ref{app:methods}--\ref{app:orbit}. 

At the distance of Haumea's two satellites, only the quadrupole-order harmonics are currently detectable \citep[][]{proudfoot2024bpm3}. Assuming Haumea's shape is symmetric under a half rotation, only two quadrupole-order harmonics are non-zero, $J_2$, the zonal harmonic describing Haumea's dynamical oblateness, and $C_{22}$, the sectoral harmonic describing Haumea's dynamical prolateness. 
Since Haumea's rotation period \citep[3.915 hr;][]{rabinowitz2006photometric} is much shorter than the satellites' orbit periods \citep[18.3 days, 49.5 days][]{rb09}, the effect of \ct{} averages out and contributes negligibly to the satellites' motion \citep{proudfoot2021prolate}. 

To quadrupole order, the \jt{} harmonic can also be expressed as a function of the principal moments of inertia of a body, 
\begin{align}
    J_2MR^2 = C-\frac{1}{2}\left( B + A \right)
\end{align}
\noindent where $M$ is Haumea's mass, $R$ is a reference radius typically taken as Haumea's volumetric radius, and $A,B,C$ are the moments of inertia along Haumea's long ($a$) axis, intermediate ($b$) axis, and short ($c$) axis. Since $A,B,C$ are functions of not only Haumea's external figure but also the density structure within Haumea, \jt{} encodes information about Haumea's shape and interior. In Appendix \ref{app:j2contributions}, we account for other potential sources of \jt{} --- like undiscovered inner satellites and rings --- and conclude \added{no plausible sources other than Haumea itself} can meaningfully contribute to our \jt{} \added{determination}. \added{We note that we use $R=772$ km \citep{proudfoot2026occ} as Haumea's reference radius throughout this work to convert our fitted parameter $J_2R^2$ into one that can be compared to the broader literature.}

Our new dynamical model for the Haumea system provides the first precise \added{determination} of Haumea's \jt, allowing us to peer inside Haumea when paired with a shape model for Haumea's external figure. \citet{ortiz2017size} reported the results from a stellar occultation by Haumea in 2017, providing the first 3D shape model of Haumea. Assuming Haumea was at light curve minimum during the occultation, they suggest $a=1161\pm30$ km, $b=852\pm4$ km, and $c=513\pm16$ km, which implies $J_2 = 0.260\pm0.013$ when assuming a homogeneous structure \citep[see][for more discussion about how to calculate \jt]{ragozzine2024beyond}.

\citet{proudfoot2026occ} iterated on this model by allowing Haumea's rotation phase to match the measured value during the occultation \added{\citep[see][for a discussion of the rotation phase during the occultation]{dunham2019haumea}}, deriving $a = 1061^{+87}_{-71}$ km, $b = 844^{+5}_{-7}$ km, and $c = 514^{+18}_{-19}$ km. Again, assuming a homogeneous density structure within Haumea, this yields $J_2 = 0.218^{+0.035}_{-0.025}$. 

Our orbit fits (see Appendix \ref{app:orbit}) provide $J_2 = 0.143^{+0.025}_{-0.024}$, rejecting the homogeneous density models at 2.1--4.2 $\sigma$, depending on the shape model used. This is strong evidence that Haumea is differentiated. To more fully explore this possibility, in the next section, we search for hydrostatic equilibrium models that match both our \jt{} measurement and the occultation-derived shape. 

\section{Haumea's Interior}

\subsection{Homogeneous models}
Before exploring differentiated interior models, we first ask a simpler question: can any homogeneous body match Haumea's observed shape, mass, and rotation period? For a homogeneous, incompressible, self-gravitating fluid in hydrostatic equilibrium (HE), the shape is uniquely determined by spin/density via the classical Jacobi ellipsoid formalism \citep{chandrasekhar1987ellipsoidal}. 

We use Haumea's $a$- and $b$-axes from the \citet{proudfoot2026occ} posterior distribution for the shape of Haumea to create a selection of trial \added{ellipsoidal} shape models. Using the known rotation period of Haumea, we then calculate the density \added{and mass} of the ellipsoidal shape. These trial \added{ellipsoidal} shapes can then be compared to Haumea's true mass (as determined by our orbit fit) and $c$-axis (as determined by occultations). We find that homogeneous HE models based on Haumea's $a,b$-axes \added{from the \citet{proudfoot2026occ} occultation-derived model} significantly overpredict Haumea's true mass and $c$-axis. Statistically, when looking at an ensemble of $10^8$ homogeneous HE models, only 0.13\% of them achieve mass, shape, and $J_2$ consistent within $3\sigma$ of our measurements, naively providing a $3\sigma$ rejection of the homogeneous HE hypothesis. Looking at the 0.13\% of models that are compatible within our arbitrary bounds, most lie near the edge of the limits, implying the true statistical significance is $>3\sigma$. 

Although TNOs as large as Haumea are typically considered to be in hydrostatic equilibrium at their current rotation periods \citep{2008Icar..195..851T}, it is possible that Haumea may retain a fossilized HE shape that was ``frozen-in'' when Haumea had a shorter rotation period. Tidal effects from Haumea's satellites could then despin Haumea to its current 3.915 hr rotation period \citep[see][and Section \ref{sec:fossil}]{2026A&A...710L...2S}. To explore this hypothesis, we evaluated a variety of rotation periods, finding that we can reject the homogeneous HE hypothesis at all rotation periods shorter than Haumea's current period with $>3\sigma$ confidence. 

In conclusion, we reject the homogeneous HE model, suggesting that Haumea's interior is more complex. This motivates comparison of more realistic two-layer models, where Haumea is differentiated into an ice-rich outer crust/mantle and an inner rock-rich core, \added{with our \jt{} determination}.

\begin{figure*}[ht!]
    \centering
    \begin{subfigure}[b]{0.47\textwidth}
        \includegraphics[height=7.75cm,trim={0mm 0mm 30mm 0mm},clip]{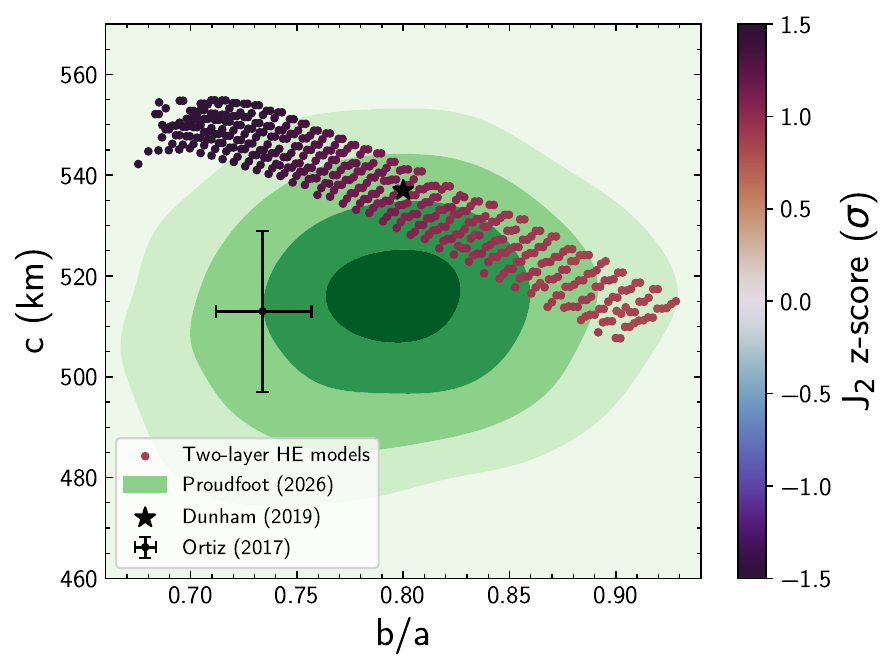}
    \end{subfigure}
    \hspace{-1.5em}
    \begin{subfigure}[b]{0.52\textwidth}
        \includegraphics[height=7.75cm,trim={2.5mm 0mm 0mm 0mm},clip]{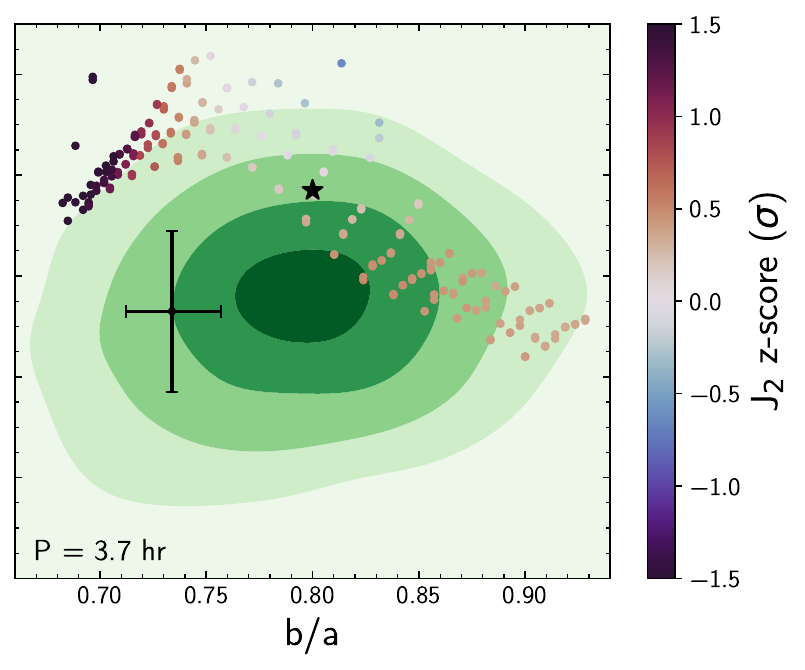}
    \end{subfigure}
    \caption{Two layer hydrostatic equilibrium models of Haumea's interior derived from \texttt{kyushu}. Green shaded regions show the 0.5, 1.0, 1.5, and 2.0 $\sigma$ contours of the occultation-derived shape constraints \citep{proudfoot2026occ}. The color of individual points show the $z$-score (i.e., how many $\sigma$ away the hydrostatic model is from our measurement) of their calculated \jt{} harmonics when compared to our \jt{} measurement from the satellite orbits. In the left panel, we show models with the nominal mass and rotation period of Haumea with mantle density of 921 kg m$^{-3}$. On the right, we show similar models, but instead for a rotation period of 3.7 hr to mimic the expected effects of a fossil-shape. }
    \label{fig:he_models}
\end{figure*}

\subsection{Two-layer models and beyond}

\citet{dunham2019haumea} (hereafter D+19) provided an initial exploration of two-layer triaxial hydrostatic equilibrium (HE) models of Haumea, showing that Haumea's occultation-derived shape, mass, and rotation period can be consistent with a body in HE. They suggested an external figure for Haumea of $a=1050$ km, $b=840$ km, and $c=537$ km with a core with axes $a_c=883$ km, $b_c=723$ km, and $c_c=470$ km. The authors found a core density of 2680 kg m$^{-3}$ --- consistent with hydrated rock --- overlaid with an icy mantle with density 921 kg m$^{-3}$. Although such a two-layer model has $J_2 = 0.1725$, just outside the $1\sigma$ uncertainties of our \jt{} measurement, comparison with the \citet{proudfoot2026occ} shape model shows that the inferred $c$-axis is also somewhat too large by $1.3\sigma$. 

The D+19 model is a clear improvement over the homogeneous models, but still does not precisely match the \jt{} and shape constraints. Since both of these constraints are independent of one another, their likelihoods multiply and yield a modestly less favorable joint fit than either constraint alone. Although the D+19 model cannot be statistically rejected, this indicates other models may better match the occultation and dynamical constraints. 

To explore such models, we used the \texttt{kyushu} code (D+19, \citealp{noviello2022let}) to find a variety of two-layer HE solutions. 
\added{\texttt{kyushu} takes input $a,b$-axes and mantle density and attempts to find a two-layer HE model that matches Haumea's mass and rotation period. The code outputs the value for Haumea's $c$-axis, the core shape, and the core density. }
The full details of how we conducted the two-layer HE modeling are in Appendix \ref{app:kyushu}. 

We looked for HE solutions with the same mantle density as that explored in D+19 (921 kg m$^{-3}$) across a range of $a$- and $b$-axes permitted by shape models. We show the results of this scan in the left panel of Figure \ref{fig:he_models} where we compare the occultation-derived shape models with the two-layer HE models. About $\sim24\%$ of these models provide a somewhat better fit to the overall shape of Haumea than the nominal D+19 model, but do not appreciably improve the match to our \jt{} measurement. 

We also tested a variety of different mantle densities, which can in principle increase/decrease the \jt{} as mass is more concentrated at Haumea's center. 
In practice, we found that within a range of likely mantle density \citep[$<1100$ kg m$^{-3}$;][]{noviello2022let} there was little improvement over the nominal value. Denser mantles provide a slightly worse match to the \jt{} measurement and shape model.

\begin{figure*}
    \centering
    \includegraphics[width=0.8\textwidth,trim={0mm 50mm 0mm 35mm},clip]{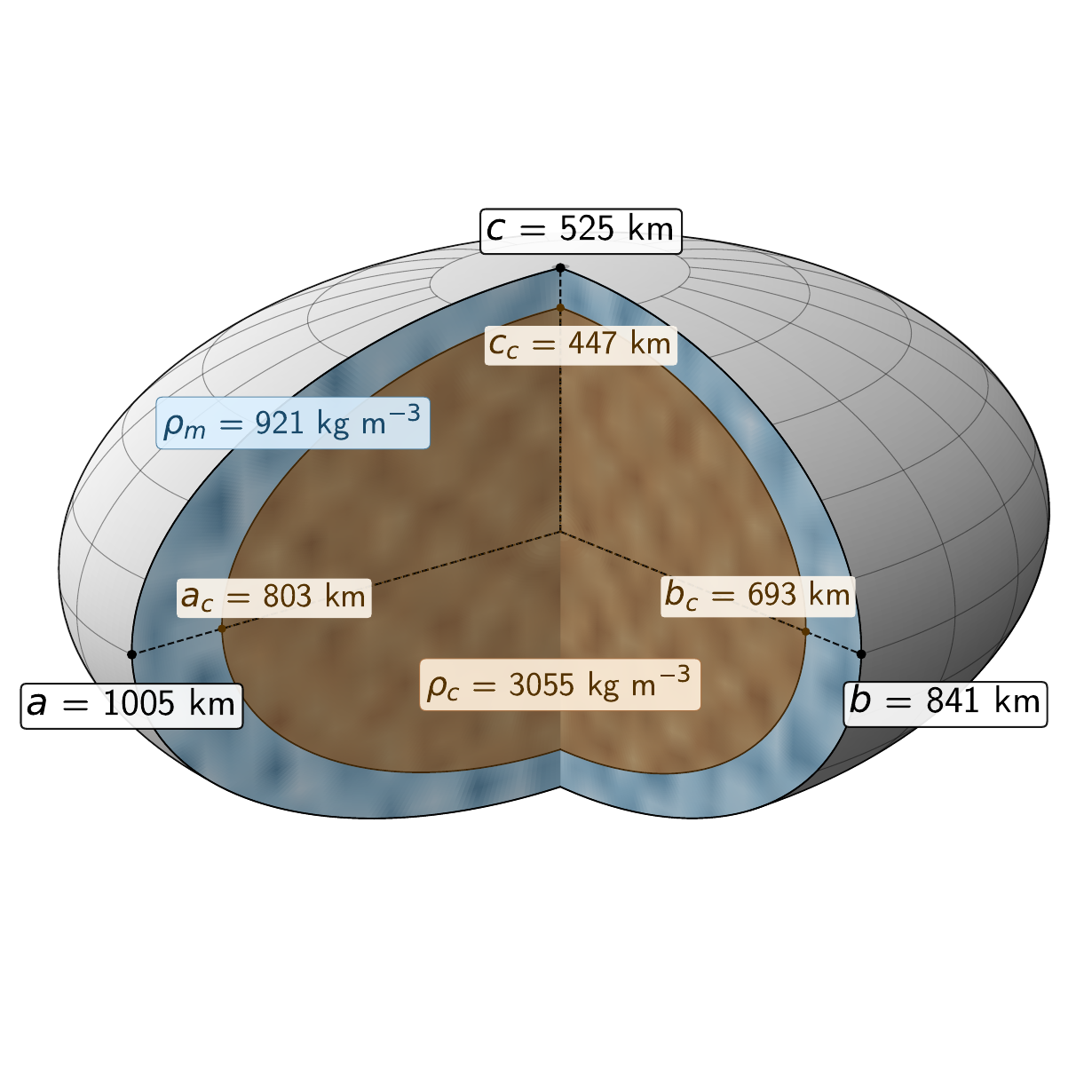}
    \caption{An example of a possible fossil HE figure that matches well the 2017 occultation shadow and the \jt{} constraint from our orbit fit. Although this model provides the closest statistical match to the shape and \jt, a variety of models with varying core density, size, and shape are possible.}
    \label{fig:interior}
\end{figure*}

We also tested the fossil-shape hypothesis with the two-layer models. We tested a rotation period of 3.7 hr, rather than doing a full scan over the possible rotation periods. \added{Although an arbitrary choice, 3.7 hours provides a good balance between overall stability against rotational mass shedding and appreciable change in the shape model outputs. We note that \citet{noviello2022let} shows that Haumea could have had rotation periods as short as 3.4 hours, although these undergo rotational shedding. }This simple modeling provides an initial exploration of the fossil-shape hypothesis and are displayed in the right panel of Figure \ref{fig:he_models}. Here, we see a group of models with closer agreement with our \jt{} measurement and a modest improvement in agreement with the shape model.

To illustrate the characteristics of the group of fossil models which match the shape and \jt, we show a schematic of one configuration in Figure \ref{fig:interior} which has $J_2=0.154$. 
We emphasize that this is just one of many models that reproduces the 2017 occultation shadow and the \jt{} measurement. At present, the available observations do not provide sufficient information to distinguish between competing interior models of Haumea, whether with the nominal rotation period or not. 

As an ensemble, all two-layer triaxial models which provide acceptable fits to both the occultation chords and \jt{} have core densities between 2500--3500 kg m$^{-3}$, consistent with both hydrated rock ($\sim2700$ kg m$^{-3}$; D+19; \citealp{noviello2022let}) or a partial mix of dehydrated and hydrated rock (where dehydrated rock has a typical density of 3800 kg m$^{-3}$; \added{consistent with} \citealp{noviello2022let}). Models with core densities on the lower end of this range tend to provide a better match to the shape model (as was found by D+19) but a worse match to the \jt{} measurement. The opposite also holds true where models with higher core density provide a better \jt{} match, but a worse shape model match. 


Beyond two-layer models, another possibility is that Haumea has three distinct layers rather than two. \citet{2026A&A...710L...2S} explored such a possibility and found that the addition of another layer provides considerable flexibility when fitting an occultation-bounded shape model. \added{That work showed individual three-layer models with $J_2=0.150-0.170$, although the lowest of these were also fossil shape models.} Future comparisons of such interior models with the external figure of Haumea and its \jt{} could show that three-layer models are indeed favored over two-layer ones, but better occultation and \jt{} measurements are required.

In any case, our work conclusively shows that Haumea's interior is differentiated, with a dense rock-rich core surrounded by a rock-depleted mantle. Improved constraints on both the external shape model and the measurement of Haumea's \jt{} may enable selection of a narrow subset of the broad range of acceptable models. To better measure the external figure, additional occultations should be observed. A large observing campaign targeted a May 2026 occultation by Haumea \citep{2026MNRAS.548ag692O}, with results forthcoming. To more precisely measure Haumea's \jt, additional observations of Hi`iaka and Namaka should be acquired. Our 2026 HST observations were designed as the minimum program required to detect differentiation at $\sim$3$\sigma$; a dedicated follow-up campaign could considerably improve the \jt{} measurement.

\section{Discussion}
\label{sec:discussion}

\subsection{Dwarf Planet Interiors}
\label{sec:interior}
A variety of indirect evidences have long suggested that dwarf planets are able to undergo full differentiation, but direct dynamical evidence of the rocky cores has so far been elusive. \added{Our work here provides dynamical confirmation of such a core within Haumea. Although the exact configuration of Haumea's interior is still degenerate, the shape and \jt{} of Haumea require that Haumea has differentiated (although possibly not completely) into a dense core and less dense mantle.} 

The existence of Haumea's core in the range of densities expected for (partially) hydrated rock suggests that conditions within Haumea were favorable for interactions between rock and liquid water. Water-rock reactions only proceed at temperatures \added{and consume water as part of the hydration process $\gtrsim273$ K \citep{2017GeCoA.212..324N}, so, to produce the large quantity of hydrated rock suggested by the density of the core, a substantial quantity of water must have been (at some point) liquid within Haumea. In principle, hydration reactions could be fed by a slow trickle of meltwater into the core, but given that hydration reactions are exothermic, hydration likely took place quickly in a self-propagating reaction front \citep[see][]{noviello2022let}.} 

\added{Such reactions could have either taken place due to hydrothermal convection at the base of a subsurface ocean, or more conservatively, in an intimately mixed rock-water matrix within the core. However, given the separation of a rock-rich phase, it seems reasonable the former took place rather than the latter. Indeed, more detailed simulations of Haumea's interior suggest that hydration took place after a sufficiently deep subsurface ocean formed \citep{noviello2022let}. Such a circulation can be driven by core cracking driven by pressurization of pore water (as suggested by \citealp{neveu2015core}, see also \citealp{noviello2022let}). Those thermodynamic models} suggest that Haumea could not maintain a subsurface ocean to the present day, but this should be explored in more detail\added{, especially in light of possible hotspots on Makemake's surface \citep{kiss2024prominent}}.

In the context of other dwarf planets, similar conditions have been suggested \added{to explain the CH$_4$ isotopic ratios} of Makemake and Eris \citep{glein2024moderate}. As such, in light of our constraints on Haumea's interior, we expect that \added{differentiation, formation of subsurface oceans, and locations for rock-water geochemical reactions are} common among 1000+ km TNOs. 

Another persistent question about the interiors of large TNOs is whether they retain crusts of undifferentiated material. \citet{desch2009thermal} found that 50+ km crusts of undifferentiated, porous material can be retained to the present, even when the subsurface undergoes dramatic geophysical/chemical evolution, due to the cold temperatures in the trans-Neptunian region. Later work, however, showed that a variety of other processes could provide enough energy to fully overturn these outer crusts \citep{neveu2015core}. One key difference between these models is the timing of the rock core hydration. Hydration reactions are exothermic, and can quickly propagate through a dehydrated core in a reaction front if conditions are suitable. Although the overall energy output of such reactions are modest (compared to the total energy budget across 4.5 Gyr) --- and can later be erased by endothermic dehydration reactions --- they can provide a short pulse of heat that propagates outwards contributing to full differentiation of the outer crust \citep{neveu2015core,noviello2022let}. 

As suggested by past theoretical work, \citep[D+19,][]{noviello2022let} and confirmed by our modeling, Haumea's core density, which is similar to hydrated rock, appears to be consistent with this process. Therefore, we expect that Haumea's crust should be completely (or almost completely) differentiated. A thick (10s of km) crust could be detectable by fitting three-layer HE models to the \jt{} and occultation constraints \citep{2026A&A...710L...2S}. Of course, at present such a task is beyond the observational data, but near-term improvements could enable this. 

Extending this to the other dwarf planets (Makemake, Eris, Pluto), we expect they too should have undergone such a transformation and may have differentiated crusts with hydrated cores. Dwarf planet candidates like Quaoar, Gonggong, Sedna, and potentially Orcus --- smaller than Haumea --- may straddle the line where this tipping point occurs. Investigation of these transition bodies using dynamical, spectroscopic, and occultation observations is a high priority.

One hint to where hydration reactions can play a role might be found in the TNO size--density relationship \citep[see][for the most up-to-date version]{proudfoot2026occ}. It appears that density plateaus around the diameter of Quaoar ($D\sim1100$ km), after which size appears far less strongly correlated with density. We speculate that if TNOs are made of similar rock-ice mixtures \citep[about 70\% rock by mass;][]{bierson2019using}, the hydration of rock predicts full interior melting and elimination of all/most pore space. Below this threshold, varying levels of partial differentiation and/or pore compaction remain \citep[see][]{bierson2019using}. Dedicated thermodynamic modeling may be able to shed further light on this hypothesis.

\subsection{Haumea's Missing CH$_4$}
\label{sec:ch4}
Haumea's confirmed \added{(probably hydrated)} rocky core invites comparison with other large TNOs like Eris and Makemake, both of which show evidence for geochemical production of CH$_4$ on their surfaces \citep{grundy2024measurement,glein2024moderate}. Quaoar, Gonggong, and Sedna --- still large in their own right but smaller than Haumea --- appear to be able to retain at least some light hydrocarbon species on their surfaces, with internal geochemical processing as a strong candidate for their source \citep{emery2024tale}. Despite Haumea's comparable size to Makemake, no such hydrocarbon inventory has been detected on its surface. 
Since water-rock reactions may fuel CH$_4$ production on Makemake and Eris \citep{glein2024moderate}, and we find evidence for water-rock reactions based on Haumea's hydrated core, where is Haumea's CH$_4$?

We consider three possibilities. First, although a hot core provides the requisite energy source for geochemistry, if the ingredients are not present, no CH$_4$ will be produced. Given Haumea's known collisional history \citep{2007Nature..446..296,proudfoot2019modeling}, it is possible that the necessary carbon-bearing species were lost during the collision \citep[see][]{2009A&A...496..547P}. Although this could explain the lack of CO and potentially CO$_2$, depending on the conditions in the immediate aftermath of the collision, refractory organics (like those present in comets; \citealp{2017MNRAS.469S.712B}) would likely remain, leaving the opportunity for thermogenic production of CH$_4$ by ``cooking'' such materials \citep[see][for a discussion of thermogenic CH$_4$ production]{glein2024moderate}. \added{Haumea's core is expected to remain hot ($>700$ K) to the present, so considerable energy for thermogenic production is possible if the refractory organics are present. As such, we disfavor the ``no carbon'' hypothesis.}

Another possibility is that the proto-Haumea had substantial CH$_4$, but geochemical activity largely ceased before the last ``reset'' of Haumea's surface, presumably during the massive collision that created the Haumea family. This reset could de-volatilize Haumea's outer layers, leaving little/no trace of the original CH$_4$ inventory. If Haumea never regained a subsurface ocean conducive to geochemical activity, a major pathway for CH$_4$ would be eliminated. This again does not account for the possibility of thermogenic CH$_4$ production from refractory organics. \added{Again, this leads us to somewhat disfavor this hypothesis.}

\added{As a slight modification of these hypotheses, it may be possible that CH$_4$ produced after Haumea's collision either may not have the ability to make it to Haumea's surface, or cannot survive long enough there to spectrally affect the surface. More detailed investigations into the expression of geochemically created species, and their subsequent lifetimes in cold traps, may provide more insight into whether these hypotheses can truly be rejected.}

One last speculative possibility is that Haumea's surface may be covered in a thin veneer of ice-rich material, while the underlying substrate could differ substantially. Since spectral observations only probe the top few $\mu$m of a body's surface, only a thin layer is required to completely hide Haumea's putative hydrocarbon inventory. In the aftermath of the creation of Haumea's family and satellite system, considerable debris likely persisted within the Haumea system. 
If one (or more) small satellites (in addition to Hi`iaka and Namaka) were eventually destroyed after Haumea's geochemical activity had slowed (or stopped), the debris could have blanketed Haumea's surface in a thin layer of family/satellite-like ice-rich material. For reference, a Namaka-mass \added{of material} could coat Haumea's surface in an uniform surface layer $\gtrsim$150 m thick depending on the density of the veneer. \added{As such, even a tiny moon which is accreted at relatively low efficiency could meaningfully alter Haumea's surface}. Although ongoing cratering of the surface might unearth buried hydrocarbons, ejecta exchange between the satellites and Haumea could maintain a relatively clean surface \citep{2009Icar..199..571S,2015Icar..246..360P}. \added{The photometric properties of Haumea's surface \citep[e.g.,][]{verbiscer2022diverse} could possibly help to understand whether this hypothesis is viable. }

No matter the origin of Haumea's missing CH$_4$, two open puzzles have the potential to lead to deep insight. First, Haumea does exhibit some compositional variability. The ``dark red spot'' (DRS) is a region of Haumea that has a lower albedo and a redder spectrum \citep{lacerda2009time}. 
Rotationally-resolved NIR spectra with JWST are one way to better understand the DRS, and could reveal traces of Haumea's hydrocarbon inventory. 

A second clue is Haumea's albedo. In contrast with the trend for large TNOs to have higher albedos than their satellites \citep{brown2018medium,2025epsc.conf..905K,2026ApJ..1002L..37F}, Haumea's albedo is significantly lower than its satellites and family \citep{elliot2010size,ortiz2017size,fernandez2025accurate,2025A&A...703A.147O}. Although the DRS likely plays a role in at least some of the albedo difference, it is unclear if these differences are due to compositional differences or grain properties. JWST observed Haumea, its satellites, and several of its family members, so archival spectral analysis of these guaranteed time observation datasets can likely begin to answer at least part of this mystery.

\subsection{Constraints on the Fossil Shape Hypothesis}
\label{sec:fossil}
Our preference for a fossil shape for Haumea, although not yet statistically significant, is worth considering in greater detail, especially in light of its possibility to provide crucial information about the formation of Haumea's family and satellites.

Such a shape requires that Haumea's lithosphere solidified before Haumea's final spin state was fully established. The classic mechanism for the formation of fossil shapes is spin down from tidal interactions. At present, both satellites are too distant to provide any meaningful tidal torques \citep{2026A&A...710L...2S}, but presumably at some point in their histories they could have been far closer to Haumea. The outward tidal migration rate of a satellite is $\dot{a}\propto a^{-11/2}$ and the despinning rate $\dot{\omega}\propto a^{-6}$ \citep{1999ssd..book.....M,boue2019tidal}, so the initial phases of tidal migration/despinning happen very rapidly followed by a long tail of extremely slow evolution \citep[see also,][]{hastings2016short}. As such, for an appreciable fossil shape to form, Haumea's crust would have had to freeze very early during the satellite's outward migration, imposing a strict timing requirement on the tidal and cooling timescales. Given the fact that we infer Haumea's full crust to have undergone differentiation, it seems unlikely that it cooled quickly enough to form the thick shell required to support the fossil shape. \added{Detailed simulations, like those done for other TNO moons \citep{2021PSJ.....2....4R,2026PSJ.....7...68E}, could be done to confirm this qualitative conclusion.}

Instead, Haumea's internal geophysical evolution could play a role. \citet{noviello2022let} explored such a hypothesis, specifically as a mechanism for creating Haumea's family. In their models, growth and changes to the hydration state of the core\added{, as well as material shedding from Haumea's $a$-axis, }provide a way to change Haumea's moment of inertia, thereby changing the rotation period. One possibility raised by this \added{scenario} is that Haumea could retain a fossil shape. \added{Timing constraints are not as clear of an issue here, given that internal changes can span over Gyr timescales. But, full 3D thermodynamic modeling is needed to fully confirm that Haumea's lithosphere could cool enough to support a fossil shape before the final changes in Haumea's interior structure. }

Confirmation of a fossil shape would provide much needed progress on the unsolved mystery of the formation of Haumea's satellites and family. If a fossil bulge does exist, it will likely place meaningful constraints on Haumea's history, providing a way to differentiate between formation models \citep[e.g.,][]{schlichting2009creation,leinhardt2010formation,ortiz2012rotational,proudfoot2022formation,noviello2022let}. 
\added{For example, if Haumea's shape implies a primordial spin significantly faster than today's, it makes family formation via destruction of satellites much less likely, as those models already struggle to explain Haumea's rapid rotation without more complex, multi-collision chains. A fossil shape also provides considerable --- but not overwhelming --- support to the family formation via geophysical evolution \citep{noviello2022let}. On the other hand, if a fossil shape is ruled out, little can be said about Haumea's past, as presumably any trace of previous epochs of different rotation rates might have been erased.}

We stress that this remains speculative. A statistically robust detection of the putative fossil figure requires, at minimum, a better shape model for Haumea's external figure and a more precise determination of \jt{} from satellite astrometry.

\section{Conclusions}

Combining two decades of HST astrometry with occultation-derived shape constraints, we present the first precise dynamical measurement of Haumea's \jt. Combined with occultation-derived shape models, \added{and consistent with previous modeling}, we find Haumea is incompatible with a homogeneous hydrostatic equilibrium figure at $>3\sigma$ confidence. Haumea has a dense, rock-rich core \added{(which is likely hydrated)} overlaid by an ice-rich mantle --- the first conclusive dynamical confirmation of differentiation of a large trans-Neptunian object. \added{Although significant uncertainties remain in the exact configuration, we suggest that Haumea could have sustained liquid water in its interior, possibly in a subsurface ocean, which underwent significant water-rock hydration reactions.}

Our present data cannot distinguish between the range of viable two-layer (and possibly three-layer) interior models, nor confirm or rule out a fossil, frozen-in shape at the current level of statistical significance. Both are within reach of near-term observations; continued astrometric tracking of Hi`iaka and Namaka will tighten \jt{} constraints, while future stellar occultations will refine Haumea's external shape. Together, these should be sufficient to narrow possible interior models and test the fossil-shape hypothesis more conclusively.

\added{Although Haumea is just one example, our dynamical confirmation of its differentiation and core hydration state can be applied more broadly to other TNOs. Haumea has a similar size and density to other 1000+ km TNOs, indicating that its interior may be broadly similar. Taken together with recent JWST findings that geochemistry may have been active in other large TNOs \citep{emery2024tale,glein2024moderate,grundy2024measurement}, we suggest that most 1000+ km TNOs differentiated and had liquids within their interior, either as liquid in the pores of a rocky core or as full subsurface oceans. Confirmation of such configurations can be had by continued observations of TNO satellites, as well as thermodynamic modeling, improved spectral observations, occultation observations, and others. Haumea too will benefit from more observations:} as with any good mystery, many of Haumea's unique traits still remain unexplained. 

\begin{acknowledgments}
We thank Yücel Kili\c{c} and Jos\'{e} Mar\'{i}a G\'{o}mez-Lim\'{o}n for productive and friendly conversations about the preliminary results of the May 4, 2026 occultation by the Haumea system. We also thank an anonymous reviewer for their comments which made this work much clearer. We also thank the BYU Office of Research Computing for their dedication to providing computing resources, without which this work would not have been possible. 

This work is based on observations with the NASA/ESA Hubble Space Telescope obtained from MAST at the Space Telescope Science Institute, which is operated by the Association of Universities for Research in Astronomy, Incorporated, under NASA contract NAS5-26555. These observations are associated with programs 10545, 10860, 11169, 11518, 12243, 13873, and 18002. Support for Program number 18002 was provided through a grant from the STScI under NASA contract NAS5-26555.

B.P. and F.L.R. are supported by the University of Central Florida Preeminent Postdoctoral Program (P$^3$). 

\end{acknowledgments}

\begin{contribution}
B.P. led the overall analysis, wrote the manuscript, and was PI of the 2026 HST program. D.R. provided access to computing resources and aided with data interpretation. W.G. analyzed all HST observations. F.L.R. and B.J.H. observed the occultation by Haumea and Namaka and were involved with writing the HST proposal. All authors contributed to interpretation of the results and editing the manuscript. 
\end{contribution}

\appendix

\begin{deluxetable*}{cccCCCCCCCC}
\tabletypesize{\footnotesize}
\tablewidth{\textwidth}
\tablecaption{Observed Astrometric Positions of Haumea's Satellites\label{tab:observations}}
\tablehead{
Julian Date & Date & Instrument & \Delta x_N & \sigma_{\Delta x_N} & \Delta y_N & \sigma_{\Delta y_N} & \Delta x_H & \sigma_{\Delta x_H} & \Delta y_H & \sigma_{\Delta y_H} \\
 & & & \textrm{('')} & \textrm{('')} & \textrm{('')} & \textrm{('')} & \textrm{('')} & \textrm{('')} & \textrm{('')} & \textrm{('')} 
}
\startdata
2460751.00318 & 2025 Mar 16 & Occultation & +0.15046 & 0.01000 & -0.14263 & 0.01000 & \nodata & \nodata & \nodata & \nodata \\
2461110.19306 & 2026 Mar 10 & HST/WFC3 & -0.19862 & 0.00200 & +0.13631 & 0.00200 & +0.48194 & 0.00200 & +1.20943 & 0.00200 \\
2461136.19373 & 2026 Apr 5 & HST/WFC3 & +0.23640 & 0.00200 & -0.04579 & 0.00200 & -0.30211 & 0.00200 & -1.40270 & 0.00200 \\
2461148.01330 & 2026 Apr 17 & HST/WFC3 & -0.30640 & 0.00200 & -0.12439 & 0.00200 & +0.54671 & 0.00200 & -0.22363 & 0.00200 \\
\enddata
\end{deluxetable*}

\section{HST Observations}
\label{app:methods}

Considerable effort has been put into observing Haumea, Hi`iaka, and Namaka since their discovery two decades ago, with seven HST programs imaging the system since early 2006. The full description of all observations acquired between 2006 and 2015 can be found in \citet{rb09} and \citet{proudfoot2024bpm3}. Here, we describe three new epochs of imaging in 2026 taken as part of Cycle 33 program 18002 (PI: B. Proudfoot). In addition, we also describe a single-chord occultation observation in 2025 that detected both Haumea and Namaka, allowing for a determination of Namaka's position relative to Haumea (see Appendix \ref{app:occult}).

Observations of the Haumea system during Cycle 33 were planned based on the orbit solution of \citet{proudfoot2024bpm3}. Using that work's orbit posterior distribution, we propagated a statistical sample of orbit solutions to get an idea of the observability (based on the Haumea-satellite and satellite-satellite separation) and uncertainty on the satellites' positions. To maximally constrain the orbits of the satellites, we identified possible observing times where Hi`iaka and Namaka's positional uncertainty were highest. With an approximately edge-on geometry, these moments typically corresponded to times when the on-sky velocity of Hi`iaka and Namaka (in the Haumea-centered frame) were highest. Three single-orbit observations were scheduled to occur at particularly constraining moments, taking care to ensure each observation captured the system at a different orbital configuration.

Each single-orbit visit with HST consisted of a straightforward sequence of visible wavelength imaging with the Wide Field Camera 3 (WFC3), which offers excellent spatial resolution to resolve the system components. Since Namaka is relatively bright when compared with other TNO satellites \citep[$V\sim22$;][]{brown2006satellites}, deep exposures were not required to achieve \added{SNR$>20$}. As such, 22 individual exposures of the system were taken each orbit using the F350LP filter\added{, which achieved SNR$\sim45$ on Namaka in each exposure, maximizing astrometric precision}. The astrometric positions of the satellites were then extracted from calibrated images using previously described PSF fitting techniques \citep[e.g.,][]{grundy2015mutual,grundy2019mutual} based on modeled PSFs from TinyTim \citep{krist201120}. 

The relative positions of Hi`iaka and Namaka during our observations are shown in Table \ref{tab:observations}. For brevity, we refer the reader to \citet{proudfoot2024bpm3} for a full table with archival HST astrometry. We use the entire HST dataset \citep[notably discarding the extensive Keck dataset based on the suggestion of][]{proudfoot2024bpm3} for this work, although we implement a noise floor of 2 mas to avoid overinterpreting small systematic noise sources in the data, similar to various recent works \citep{collyer2025synchronous,proudfoot2025beyond}. 

All new HST observations are publicly available at DOI:\dataset[10.17909/qwfv-qj17]{http://dx.doi.org/10.17909/qwfv-qj17}.

\begin{figure*}
    \centering\includegraphics[width=0.99\textwidth]{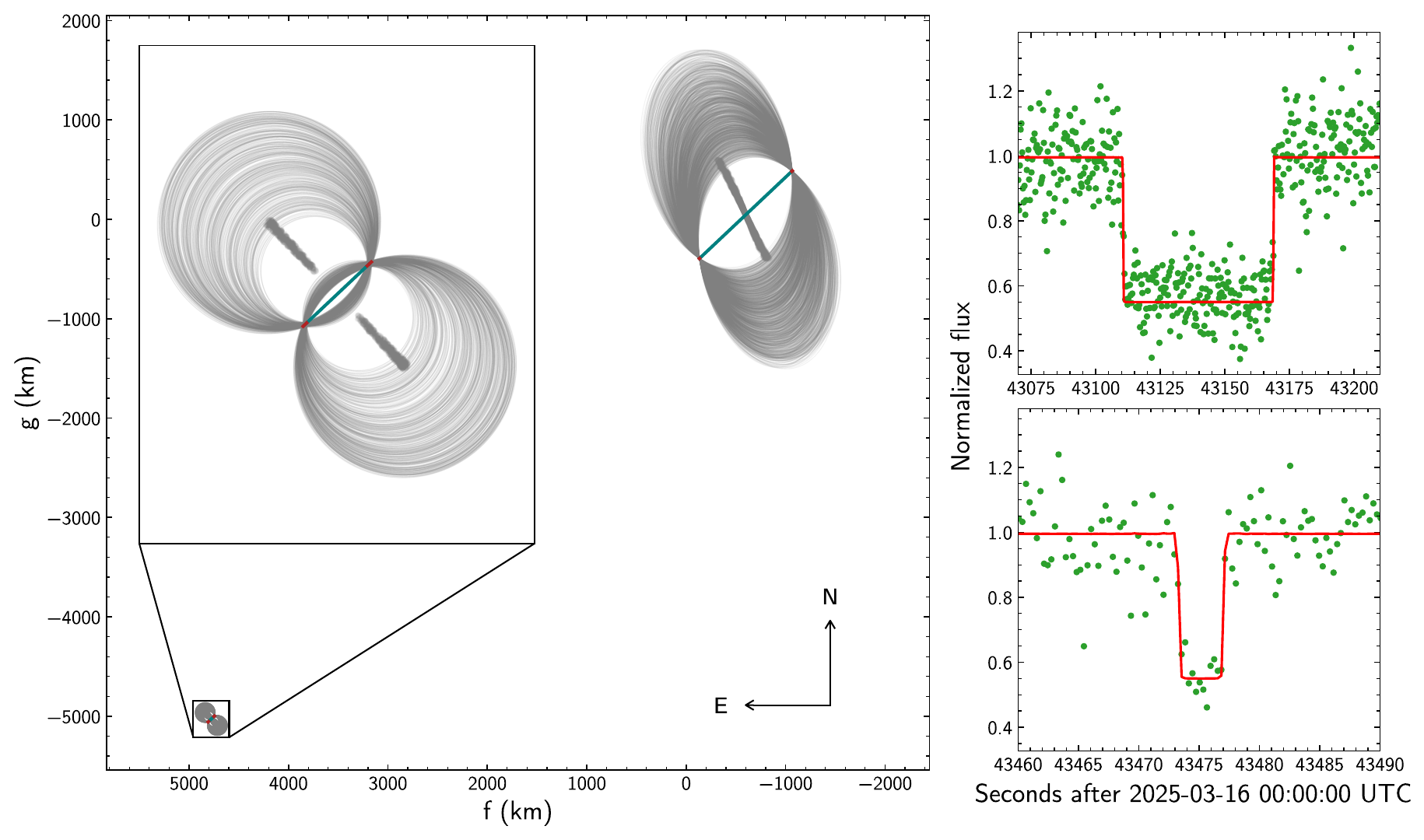}
    \caption{Left: The single chord detection of both Haumea and Namaka projected onto the sky plane (with origin defined by the JPL\#125 ephemeris of Haumea). Ellipses and circles show the possible limb solutions for Haumea and Namaka (see inset). Right: Occultation light curves showing the detections of Haumea and Namaka. Red lines show the modeled light curve used to extract immersion and emersion times. }
    \label{fig:occultation}
\end{figure*}

\section{Haumea-Namaka Occultation}
\label{app:occult}
In addition to the HST dataset, we also include a unique double occultation observation by Haumea and Namaka. A preliminary description of the detections was reported by \citet{2025RNAAS...9...62R}. The pipeline used to predict the occultation is described in \citet{Rommel_2026}. With just a single chord, the precision of the measurement is somewhat uncertain at $\sim10$ mas, but its inclusion does further refine our orbital model. 

Observations of a predicted occultation by both Haumea and Namaka were conducted from NASA's Infrared Telescope Facility (IRTF) on 2025-03-16 UT. Full details of the observations are given in \citet{2025RNAAS...9...62R}. With a normalized light curve, we fit for the times of immersion (i.e., the occulted star disappearing) and emersion (i.e., the occulted star reappearing) using the SORA package \citep{2022MNRAS.511.1167G}, which were then projected onto the sky-plane ($f$, $g$ referenced to the JPL \#125 ephemeris of Haumea). We show both the light curves and the projected chords in Figure \ref{fig:occultation}. To extract relative astrometry of Namaka, a full elliptical fit to occultation chords is typically required, allowing for a precise measurement of each body's position. Unfortunately, with just a single chord, a full shape cannot be fit, so instead we use the known shapes of Haumea and the assumed size of Namaka. 

For Haumea, we used the \citet{proudfoot2026occ} shape model propagated to the time of the 2025 occultation. We used a rotation phase of 32.1$\degr$ based on the well-studied light curve of Haumea (J.L. Ortiz, pers. comm.) with an uncertainty of 5$\degr$. The propagated shape model predicted that at the time of the 2025 occultation the apparent semi-major axis of the elliptical limb would be $a'=1006\pm67$ km, the apparent semi-minor axis of the elliptical limb would be $b'=572\pm15$ km, and a position angle of $\theta=103.3\pm1.6\degr$. Using this predicted shape and its uncertainty, we used the SORA \texttt{fit_ellipse} function to explore the range of $f,g$ that the single-chord was compatible with, finding $f=-591\pm122$ km and $g=62\pm255$ km. 

For Namaka, we followed a similar process, although we instead assumed a circular limb profile for Namaka, assuming that Namaka has a diameter of $150\pm50$ km \citep{muller2019haumea}. Given this, two possible families of solutions exist, one with Namaka's center to the north of our chord and another to the south. To provide a single measurement, we averaged these two families of solutions together, giving $f=4778\pm46$ km and $g=-5027\pm49$ km. These positions were then converted into relative astrometric offsets using the known geocentric distance to the system at the time of the occultation, providing the astrometric measurement shown in Table \ref{tab:observations}. Since the observations were only single-chords, we conservatively place a 10 mas noise floor on this measurement to account for systematic errors in both objects' shape models. 

\section{Orbit Fitting}
\label{app:orbit}

Orbit fitting was conducted using \multimoon, which is able to model both N-body and quadrupole gravitational harmonics using a Markov chain Monte Carlo (MCMC) framework \citep{ragozzine2024beyond}. Our fits used an identical set up to those used in \citet{proudfoot2024bpm3}, independently fitting the masses of all three bodies, each satellites' orbital elements, the orientation of Haumea's gravitational field, and Haumea's \jt. In total, our model has 18 free parameters. The \multimoon{} fits we present here were conducted similarly to past works in terms of priors, walker initialization, etc. Our final MCMC run used 960 walkers which underwent a 6000 step burn-in and $10^4$ step sample. 

\begin{figure*}
    \centering\includegraphics[width=0.99\textwidth]{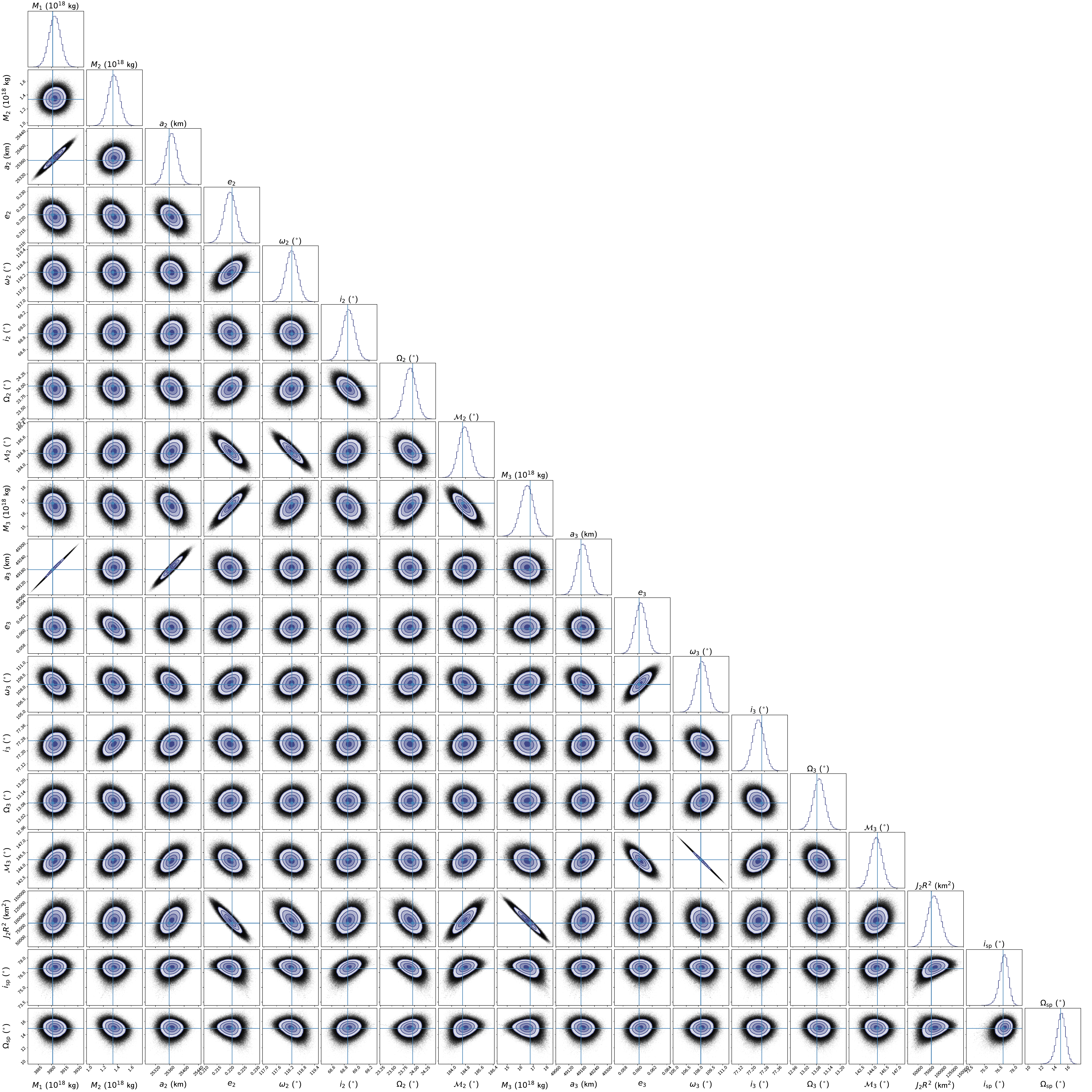}
    \caption{The 18-dimensional posterior distribution from our orbit fitting. One-dimensional histograms at the top of each column show the marginal posterior distribution for each parameter. Joint posteriors for every pair of parameters are shown beneath. Contours on the joint distributions show the 0.5-, 1-, 1.5-, and 2-$\sigma$ confidence intervals. Horizontal and vertical lines show the location of the best-fit parameter set.}
    \label{fig:corner}
\end{figure*}

The posterior distribution found from our orbit fits, as well as various derived quantities, is shown in Table \ref{tab:orbitfit}. We also display the 18-dimensional posterior distribution as a corner plot in Figure \ref{fig:corner}. The best fit parameter set in our MCMC chain achieved a $\chi^2=198$ with 104 degrees-of-freedom, giving a reduced-$\chi^2 = 1.9$. Although somewhat elevated, given that our fit combines 20 years of data from four different HST instruments using a variety of filters and instrument settings, small systematic data cross-calibration issues are not unexpected. 

Furthermore, using Haumea as a astrometric reference may introduce systematic noise to the orbit fits. With Haumea's significantly nonspherical shape, it may be expected that, depending on the surface scattering properties, Haumea's photocenter (what is measured in PSF fitting) may not align with its center-of-mass (what is relevant for orbit fitting). On top of that, Haumea has significant albedo variegations across its surface \citep{lacerda2009time}. Haumea's longest axis covers about 1.5 WFC3 pixels, so even subtle shifts in Haumea's photocenter can contribute to a somewhat noisy fit. Similar effects have been noted in a variety of TNO satellite orbit fitting efforts \citep{buie2012orbit,collyer2025synchronous,proudfoot2025beyond}. Such effects provide statistical noise to our measurements at the level of a few mas, but they cannot mimic the substantial precession that we detect which shifts the satellites' positions by 10s--100s of mas. 

Despite these effects providing some noise to our fitting efforts, residuals are generally quite small, typically $\lesssim10$ mas. Given the WFC3 pixel scale of 40 mas px$^{-1}$, our orbit solution still provides a sub-pixel accuracy model of the system. Assuming that the non-ideal $\chi^2$ is due to unaccounted for small systematic issues, we suggest that the astrometric uncertainties in our orbit fit are underestimated by a factor of $\sqrt{198/104}\approx1.38$. This yields typical uncertainties of $\sim7$ mas for Namaka and $\sim4$ mas for Hi`iaka. 

With a reasonably small estimate for the systematic errors present in our data set, as well as a plausible explanation for the source of these effects, we consider our fits to provide a high-fidelity model of the system's orbital configuration. As further evidence for the quality of our orbit fits, we note that our model successfully produces a post-diction for the occultation by one of Haumea's satellites during a recent stellar occultation (Y. Kili\c{c}, in preparation). 

One notable aspect of our orbit fits is the orbit pole we derive based purely on dynamical constraints. Our orbit fits provide a measurement of $(\alpha_{\rm pole}, \delta_{\rm pole}) = (284.8^{+0.6}_{-0.7} \degr, -9.7^{+0.5}_{-0.4} \degr)$. \citet{ortiz2017size} discovered Haumea's ring system, which should lie in Haumea's equatorial plane \citep{marzari2020ring}, to have a pole of $(\alpha_p, \delta_p) = (285.1\pm0.5 \degr,-10.6\pm1.2 \degr)$. The excellent agreement between these two independently-measured values is a strong indication that our orbit fits provide a realistic description of the system's dynamical state. 

Another notable aspect is our new precise measurement of Hi`iaka's mass. Previous efforts to determine Hi`iaka's mass were hampered by a degeneracy between its mass and Haumea's \jt{} \citep{proudfoot2024bpm3}. During a 2021 stellar occultation, \citet{fernandez2025accurate} was able to measure Hi`iaka's volumetric diameter to be $370\pm20$ km allowing for a calculation of its density. With our updated mass measurement, we find $\rho=620\pm100$ kg m$^{-3}$. This is consistent with formation from gentle accretion in an ice-rich disk surrounding Haumea. Similar constraints could be derived for Namaka with a multi-chord stellar occultation. 

\section{Ring and Undiscovered Satellite Contributions to $J_2$}
\label{app:j2contributions}
Haumea is responsible for the bulk of the total \jt{} moment felt by the satellites, but other sources can potentially add to this. The most obvious is Haumea's ring. 

The \jt{} of the ring surrounding Haumea is given by
\begin{equation}\label{eqn:j2_ring}
    J_{2} \approx \frac{\pi r W \Sigma}{M_H} \left(\frac{r}{R}\right)^2
\end{equation}
\noindent where $r$ is the ring radius \citep[2287 km;][]{ortiz2017size}, $W$ is the ring width \citep[70 km;][]{ortiz2017size}, and $\Sigma$ is the ring's surface density. We assume a similar $\Sigma$ to that of Saturn's A ring \citep[$\sim$500-1000 kg m$^{-2}$][]{sicardy2020dynamics}. This yields a total \jt{} contribution $\lesssim10^{-6}$. As such, Haumea's ring contributes negligibly to the measured \jt. Even a much larger, denser ring could not appreciably contribute to our \jt{} measurement.

Another possible source is small satellites that remain undiscovered between Haumea and Namaka. The recent discovery of a small, interior satellite around Quaoar show that this is indeed a possibility \citep{nolthenius2025discovery,2025ApJ...993L..38P,2026ApJ...999L..39B}. Explicit searches for past satellites have been conducted previously, ruling out satellites with diameters $>60$ km within 10,000 km of Haumea \citep{burkhart2016deep}. Taking this upper limit as an illustrative case ($D=60$ km, $a=$10,000 km, $\rho=1000$ kg m$^{-3}$), we find $J_2=0.0024$, again contributing essentially no signature to the \jt{} felt by Namaka and Hi`iaka. In principle, multiple smaller satellites could exist, but a dozen or more satellites at the upper limit of past searches are required to produce even $J_2=0.01$. As such, undiscovered satellites likely contribute negligibly to \jt.

Since neither rings or undiscovered satellites provide significant \jt{} signature, we safely assume that our \jt{} measurement is attributable in its entirety to Haumea. Even if this assumption were not true, it would decrease the inferred value of Haumea's \jt, further increasing the statistical confidence on our inference that Haumea is differentiated.

\section{Two-layer Hydrostatic Equilibrium Modeling}
\label{app:kyushu}
To explore other two-layer HE models, we used the \texttt{kyushu} code, which uses the \citet{1986ApJS...61..479H,1986ApJS...62..461H} algorithm to iteratively find hydrostatic equilibrium figures for a two-layer Haumea (D+19). 

We modified the publicly availably \texttt{kyushu} code, which is written in Fortran, to use our updated mass of Haumea while also adding the ability to modify certain input parameters like the density of the icy crust (to emulate a partially undifferentiated or porous crust), the rotation period of Haumea (to explore the effects of possible fossil figures), and grid resolution, among others. Our modifications also focused on performance improvements; with only small tweaks to loop structure to reduce repeated computation, we reduced the wall-clock time required to find a single HE model by over 100$\times$ to $\lesssim90$ seconds. Lastly, we implemented a Python wrapper around the Fortran functions, allowing for automated parameter scans over inputs of interest. These modifications were benchmarked against the unmodified \texttt{kyushu} code (using identical resolution and inputs) and found agreement within floating point tolerance. We also benchmarked our version against published \texttt{kyushu} HE models \citep[see, D+19][]{noviello2022let}, finding agreement within the numerical resolution of the code.

Since a wide range of $b/a$ axis ratios are allowed by the currently available occultations by Haumea \citep{proudfoot2026occ}, we first searched for two-layer HE models across a \added{$41\times14=574$ grid across} $a\in[920,1260]$ km and $b\in[828,854]$ km assuming a crust density of 921 kg m$^{-3}$ and nominal values of Haumea's mass and rotation period. This covers the full range of $b/a$ size ratios suggested by all studies of Haumea's shape \citep{rabinowitz2006photometric,2008AJ....135.1749L,2010A&A...518L.147L,2014EM&P..111..127L,ortiz2017size,2018MNRAS.478.3159K,dunham2019haumea,noviello2022let,proudfoot2026occ}. Our full results are shown in the left panel of Figure \ref{fig:he_models}. 

Haumea could also retain a hydrostatic equilibrium figure from a prior epoch that was ``frozen-in'' after cooling in the aftermath of its formation. To test this hypothesis, we looked for two-layer HE models with a rotation period of 3.7 hours across the same $a$--$b$ grid, although with looser spacing (see right panel of Figure \ref{fig:he_models}). 

\bibliography{all}{}
\bibliographystyle{aasjournalv7}

\end{document}